\documentclass[runningheads]{llncs}
\usepackage{graphicx}
\usepackage{cases}

\usepackage{ucs}
\usepackage[utf8x]{inputenc}
\usepackage{amssymb}

\begin{document}
\title{Compartment-specific estimation of T2 and T2* with diffusion-PEPTIDE MRI}
%
%
\author{Ting Gong\inst{1}\orcidID{0000-0002-1378-8509} \and
Merlin J. Fair\inst{2}\orcidID{0000-0002-9100-3105} \and
Kawin Setsompop\inst{2,3}\orcidID{0000-0003-0455-7634} \and Hui Zhang\inst{1}\orcidID{0000-0002-5426-2140}}
\authorrunning{T. Gong et al.}
%
\institute{Centre for Medical Image Computing \& Department of Computer Science, University College London, London, UK 
\and
Radiological Sciences Laboratory, Department of Radiology, Stanford University, Stanford, CA, USA \and Department of Electrical Engineering, Stanford University, Stanford, CA, USA
}
\maketitle              
\begin{abstract}
We present a microstructure imaging technique for estimating compartment-specific T2 and T2* simultaneously in the human brain. Microstructure imaging with diffusion MRI (dMRI) has enabled the modelling of intra-neurite and extra-neurite diffusion signals separately allowing for the estimation of compartment-specific tissue properties. These compartment-specific properties have been widely used in clinical studies. However, conventional dMRI cannot disentangle differences in relaxations between tissue compartments, causing biased estimates of diffusion measures which also change with TE. To solve the problem, combined relaxometry-diffusion imaging methods have been developed in recent years, providing compartmental T2-diffusion or T2*-diffusion imaging respectively, but not T2 and T2* together. As they provide complementary information, a technique that can estimate both jointly with diffusion is appealing to neuroimaging studies. The aim of this work is to develop a method to map compartmental T2-T2*-diffusion simultaneously. Using an advanced MRI acquisition called diffusion-PEPTIDE, a novel microstructure model is proposed and a multi-step fitting method is developed to estimate parameters of interest. We demonstrate for the first time that compartmental T2, T2* can be estimated simultaneously from \textit{in vivo} data. we further show the accuracy and precision of parameter estimation with simulation. 
\end{abstract}

\section{Introduction}
Diffusion MRI (dMRI) is a unique tool for imaging tissue microstructure \textit{in vivo}. Being sensitive to neurite morphology, dMRI allows for separate modelling of intra-neurite and extra-neurite diffusion signals. This enables the extraction of compartment-specific tissue microstructure parameters useful in neuroimaging studies for quantification of healthy and pathological alterations~\cite{Alexander2019ImagingApplications}.

Despite its unique utility, characterising tissue microstructure using conventional dMRI has one significant limitation: Differences in relaxation constants of different tissue compartments are not considered explicitly. As a result, when such differences exist, one cannot disentangle differences in diffusion from relaxation between compartments. This has been suggested to cause overestimated intra-axonal fraction in the tissue fraction, as T2 of intra-axonal compartment is shown to be higher than that of extra-axonal space~\cite{Gong2020MTE-NODDI:Times}. Furthermore, the extent of this overestimation depends on acquisition settings, such as echo times (TE). This renders studies with different settings potentially incomparable and limits the interpretability of studies where alteration of both diffusion  and relaxation properties are known to co-exist, such as in brain development.

To address this limitation, combined diffusion-relaxometry techniques have been developed in recent years, which consider diffusion and relaxometry jointly in both acquisition and modelling. These methods offer two key benefits, as combined T2-diffusion methods demonstrate~\cite{Gong2020MTE-NODDI:Times,Veraart2018TETimes,Lampinen2019SearchingModeling,Kim2020MultidimensionalApplications}. On the one hand they enable the estimation of non-relaxation weighted, hence TE-independent, diffusion parameters. On the other, they allow for compartment-specific T2 relaxation times to be estimated, which is challenging using relaxometry alone.

Current combined T2-diffusion approaches are limited because they cannot provide information about T2* which offers complementary information to T2. For example, T2 and T2* have been suggested to be associated with different pools of iron distribution~\cite{Brammerloh2021MeasuringRelaxometry}; the additional information T2* provides is important in applications such as characterising multiple sclerosis lesions~\cite{Herrmann2021SimultaneousRARE-EPI}. A few combined T2*-diffusion techniques have emerged recently~\cite{Hutter2018IntegratedZEBRA,Kleban2020StrongBrain,Slator2019CombinedPlacenta}. However, they forgo the ability to estimate T2. Thus, techniques that can simultaneously map T2-T2*-diffusion, while evidently appealing, are currently unavailable.

A recently developed novel sequence called diffusion-PEPTIDE~\cite{Fair2021Diffusion-PEPTIDE:Capabilities} provides us with a unique opportunity to address this challenge. Diffusion-PEPTIDE provides rich diffusion-relaxation data, where diffusion weighted images with different T2 and T2* weightings are time-resolved across its ~50 ms readout at ~1 ms increment.  Leveraging this rich data, we propose a relaxometry-diffusion combined multi-compartment model of microstructure that captures signals from intra-neurite, extra-neurite and free water compartments. To enable a robust estimation of diffusion and relaxation parameters, a multi-stage fitting approach is developed. The approach is demonstrated with the NODDI model of tissue microstructure, yielding mapping of intra-neurite, extra-neurite T2/T2*, intra-neurite fraction, free water fraction and orientation dispersion simultaneously for the first time in human brain.

\section{Method}
This section describes the theory for determining compartmental T2, T2* and associated non-relaxation-weighted compartment fractions from diffusion-PEPTIDE measurements. We first introduce the time-resolved signal from diffusion-PEPTIDE, illustrated with a simple one-compartment model. We then define the proposed microstructure model which explains the relationship between diffusion-PEPTIDE signals and model parameters. Finally, we describe the developed multi-stage fitting approach for parameter estimation. 

\subsection{Diffusion-PEPTIDE signals}
The time-resolved signal that diffusion-PEPTIDE provides is illustrated with a simple one-compartment model. Given a diffusion weighting $\vec{b} = b\hat{n}$, the signal at time point $t$ can be expressed as:
\begin{equation}
\label{eqn:signal_single_compartment}
    S(\vec{b}; t) = S^0 E(t) A(\vec{b})
\end{equation}
where $S^0=S(b=0;t=0)$ is the initial b=0 signal, $A(\vec{b})$ is the signal attenuation factor due to diffusion weighting, which equals to 1 when b=0; E(t) is the signal attenuation factor due to transverse-relaxation weighting.

$E(t)$ distinguishes diffusion-PEPTIDE from conventional single-shot EPI-based diffusion acquisition. E(t) is expressed as~\cite{Fair2021Diffusion-PEPTIDE:Capabilities}:
\begin{subnumcases}{E(t)=}
        \exp{(-t R_2 - (\mbox{TE}_{SE} - t)R_2^{'})}, & \(\mbox{TE}_{SE}/2 < t < \mbox{TE}_{SE}\) \label{eqn:TE_dependence_large} \\
        \exp{(-t R_2 - (t - \mbox{TE}_{SE})R_2^{'})}, & \(t >= \mbox{TE}_{SE}\)
        \label{eqn:TE_dependence_small}
\end{subnumcases}
where \(R_2\) and \(R_2^{'}\) are the relaxation rates, and \(\mbox{TE}_{SE}\) is the echo time of the spin echo (SE).

Conventional single-shot EPI-based diffusion acquires data only for the time point $\mbox{TE}_{SE}$. In contrast, diffusion-PEPTIDE acquires data for a series of time points around $\mbox{TE}_{SE}$, capturing a dense sample of varying  $T_2 = 1/R_2$ and $T_2^* = 1/(R_2+R_2^{'})$ contrasts. This opens the possibility for simultaneous mapping of T2-T2*-diffusion that we will exploit in this paper.

\subsection{Forward model}
The proposed model is comprised of three non-exchanging compartments including the intra-neurite and extra-neurite spaces, to model tissue microstructure, and a free water space to account for CSF contamination. The diffusion-PEPTIDE signal \(S(b,t)\) is represented as a combination of signals from all compartments, and can be expressed mathematically as:  
\begin{equation} \label{eqn:signal_three_compartments}
S(\vec{b};t)=S_{in}^0E_{in}(t)A_{in}(\vec{b})+S_{en}^0E_{en}(t)A_{en}(\vec{b})+S_{iso}^0E_{iso}(t)A_{iso}(\vec{b})
\end{equation}
where \(S_{in}^0\), \(S_{en}^0\) and \(S_{iso}^0\) are the initial b=0 signals for the intra-neurite, extra-neurite and free water compartments respectively; \(E_{in}(t)\), \(E_{en}(t)\) and \(E_{iso}(t)\) are the corresponding attenuation factors of the transverse relaxation weighting; \(A_{in}(\vec{b})\), \(A_{en}(\vec{b})\) and \(A_{iso}(\vec{b})\) are the corresponding attenuation factors of diffusion weighting. Henceforth, $t$ and $\vec{b}$ dependencies are suppressed for clarity.

Normalising the signal to the total initial b=0 signal $S^0 = S_{in}^0 + S_{en}^0 + S_{iso}^0$, the forward model can be written in terms of compartmental signal fractions as:
\begin{equation}
\label{eqn:normalised_signal_three_compartments}
    S(\vec{b},t)/S^0 = (1-f_{iso}^0)(f_{in}^0E_{in}A_{in} + (1-f_{in}^0)E_{en}A_{en}) + f_{iso}^0E_{iso}A_{iso}
\end{equation}
where \(f_{in}^0\) and \(f_{iso}^0\) are the non-relaxation weighted intra-neurite and free water fractions defined with regard to the initial tissue and total b=0 signals respectively as \(f_{in}^0=S_{in}^0/(S_{in}^0+S_{en}^0 )\) and \(f_{iso}^0=S_{iso}^0/(S_{in}^0+S_{en}^0+S_{iso}^0 )\); \(f_{in}^0\) and \(f_{iso}^0\) are TE-independent, in contrast to the compartment fractions derived with conventional diffusion method alone, which are relaxation-weighted and hence TE-dependent as \(f_{in}(t)=S_{in}^0E_{in}(t)/(S_{in}^0E_{in}(t)+ S_{en}^0E_{en}(t))\) and \(f_{iso} (t)=S_{iso}^0E_{iso}(t)/(S_{in}^0E_{in} (t)+S_{en}^0E_{en}(t)+S_{iso}^0E_{iso}(t))\).

The attenuation factors due to transverse relaxation weighting (\(E_{in}\),\(E_{en}\)  and \(E_{iso}\)) are determined by the compartmental relaxation rates as in Eqn~(\ref{eqn:TE_dependence_large}) and~(\ref{eqn:TE_dependence_small}), which are  \(R_{2}^{in}\), \(R_{2}^{'in}\), \(R_{2}^{en}\), \(R_{2}^{'en}\) and \(R_{2}^{iso}\), \(R_{2}^{'iso}\) respectively. The attenuation factors due to diffusion weighting can be modelled in different ways~\cite{Alexander2019ImagingApplications}. Here, we use the NODDI model~\cite{Zhang2012NODDI:Brain} as an example, which models $A_{in}$ as orientation dispersed sticks with a Watson distribution, and $A_{en}$ and $A_{iso}$ as anisotropic and isotropic Gaussian diffusion respectively.

The full list of the parameters in the forward model hence include \(f_{in}^0\), \(f_{iso}^0\), \(S^0\), \(R_2^{in}\), \(R_2^{'in}\), \(R_2^{en}\), \(R_2^{'en}\), \(R_2^{iso}\), \(R_2^{'iso}\) and the other NODDI parameters.

\subsection{Multi-stage fitting}
To estimate the compartmental relaxation rates and signal fractions from the model, a multi-stage approach similar to MTE-NODDI~\cite{Gong2020MTE-NODDI:Times} is developed to alleviate the difficulty of direct optimisation of the forward model which is known to be challenging given its high-dimensional parameter space. 

\textbf{Stage 1} estimates \(f_{in} (t)\) and \(f_{iso} (t)\) at each time point with conventional diffusion method. For the chosen NODDI example, these are estimated by fitting diffusion-PEPTIDE signals at each TE with NODDI MATLAB Toolbox.

\textbf{Stage 2} takes the output from Stage 1 and the b=0 signals as input to extract parameters of interest following three sequential fittings as detailed below. 

Firstly, the intra-neurite relaxation rates can be estimated by exploiting the TE-dependence of non-diffusion-weighted intra-neurite signals, as they can be expressed mathematically by combining the estimated $f_{in} (t)$ and $f_{iso} (t)$ with the b=0 signals $S(b=0,t)$ as:
\begin{equation}
\label{eqn:stage2_intra_neurite_b0_signal}
    S(b=0,t)f_{in}(t)(1-f_{iso}(t))=S_{in}^0 E_{in}(t)
\end{equation}
This allows the estimation of its three unknowns: $R_2^{in}$, $R_2^{'in}$ and $S_{in}^0$. 

Secondly, similar to MTE-NODDI, the conventional relaxation-weighted intra-neurite fraction \(f_{in}(t)\) can be written in terms of non-relaxation-weighted \(f_{in}^0\) as:
\begin{equation}
\label{eqn:stage2_intra_neurite_fraction}
    f_{in}(t)=f_{in}^0E_{in}(t)/(f_{in}^0E_{in}(t)+(1-f_{in}^0)E_{en}(t)) 
\end{equation}
Given the estimated $R_2^{in}$ and $R_2^{'in}$ from the previous step, this allows the estimation of its three remaining unknowns: $f_{in}^0$, $R_2^{en}$ and $R_2^{'en}$.

Lastly, the relaxation-weighted free water fraction can also be written in terms of non-relaxation-weighted $f_{iso}^0$ by making using of intra-neurite signal fractions as:
\begin{equation}
\label{eqn:stage2_free_water_fraction}
    f_{iso}(t)=f_{iso}^0E_{iso}(t)/(f_{iso}^0 E_{iso}(t)+(1-f_{iso}^0 )f_{in}^0E_{in} (t)/f_{in}(t))
\end{equation}
Given the estimated $f_{in}^0$, $R_2^{in}$ and $R_2^{'in}$ from the previous two steps, the remaining unknowns ($f_{iso}^0$, $R_2^{iso}$ and $R_2^{'iso}$) can be determined.

Compartmental T2 and T2* can then be calculated from R2 and R2’ according to $T_2=1/R_2$ and $T_2^*=1/(R_2+R_2')$. As the TE range typically used in diffusion-PEPTIDE is not sensitive to the long T2 of free water, we focus mainly on the resulting T2 and T2* within tissue compartment. 

\subsubsection{Implementation Details} This subsection describes the optimisation parameter settings used in the fitting steps of Stage 2. All parameters were estimated with non-linear least squares with randomly initialised starting points. To allow robust fitting, logarithm was taken of both sides of Eqn~(\ref{eqn:stage2_intra_neurite_b0_signal}) to transform the distribution of signals, and $R_2^{in}$, $R_2^{'in}$ and $\ln(S_{in}^0)$ were estimated with starting point sampled from [1/200, 1/20] $\mbox{ms}^{-1}$, [1/2000, 1/20] $\mbox{ms}^{-1}$, $[\ln(0.01),\ln(10)]$; the $f_{in}^0$, $R_2^{en}$ and $R_2^{'en}$ were then determined from Eqn~(\ref{eqn:stage2_intra_neurite_fraction}) with starting point sampled from [0,1], [1/200, 1/20] $\mbox{ms}^{-1}$, [1/2000, 1/20] $\mbox{ms}^{-1}$; $f_{iso}^0$, $R_2^{iso}$ and $R_2^{'iso}$ were finally determined from Eqn~(\ref{eqn:stage2_free_water_fraction}) with starting point sampled from [0, 1], [1/2000, 1/20] $\mbox{ms}^{-1}$, [1/2000, 1/20] $\mbox{ms}^{-1}$. The ranges of $R_2$ were chosen to be in accordance with, or wider than, those in the literature. For all fittings, the optimisation terminates when changes to residual sum of squares between iterations are below $10^{-15}$ or when the maximum iteration of 1500 is reached. The optimization algorithm chosen for non-linear least squares uses a trust region reflective approach~\cite{Coleman1996AnBounds}; lower and upper bounds for the parameters were set to their respective initialisation ranges. 

\begin{figure}[ht]
\includegraphics[width=\textwidth]{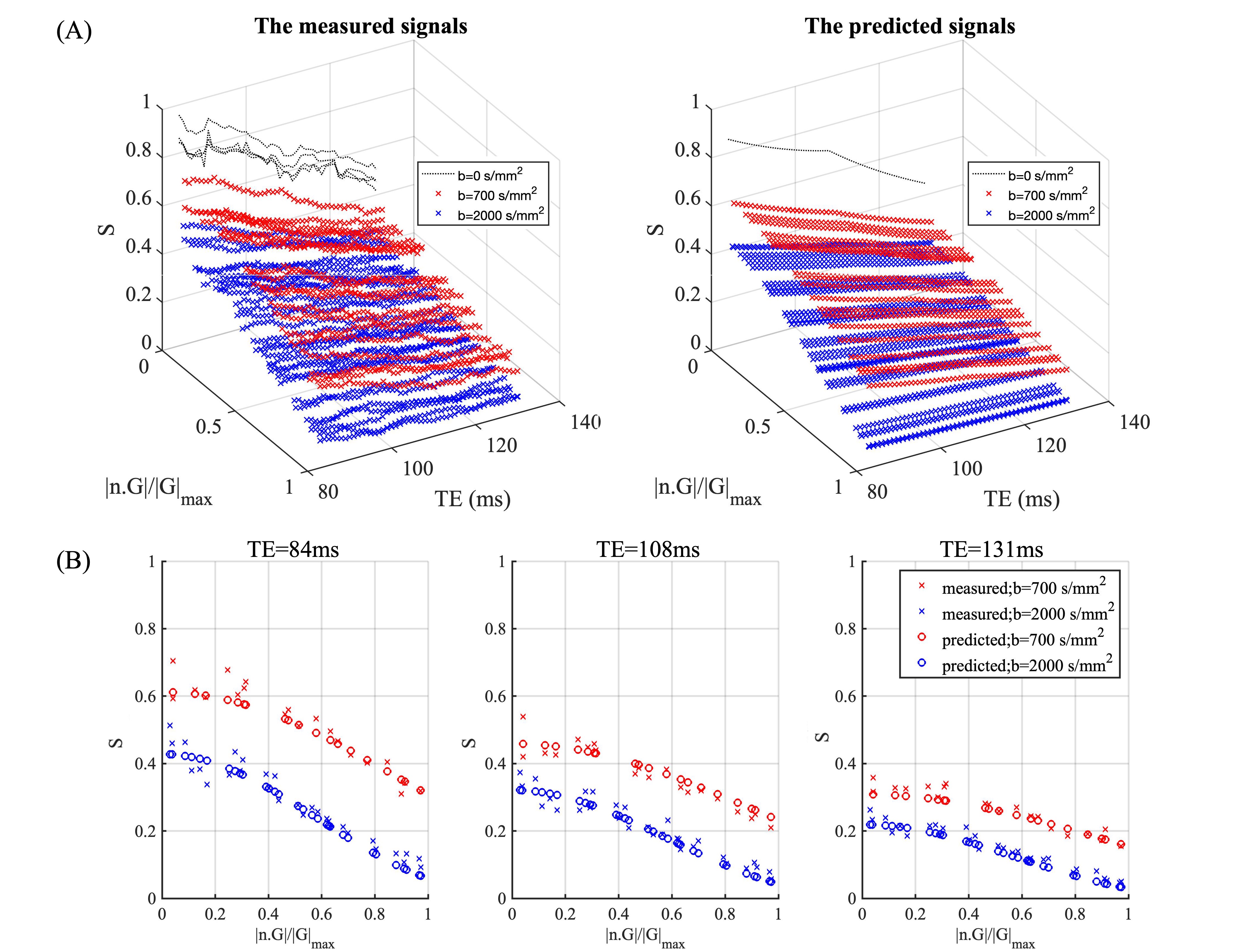}
\caption{(A) The measured and model-predicted signals from a typical WM voxel. Signals are plotted against the TE and absolute dot product between the diffusion gradient and fibre orientation for each b value. (B) The measured and predicted signals at several example TEs.} \label{In-vivo signal}
\end{figure}

\begin{figure}[h]
\includegraphics[width=\textwidth]{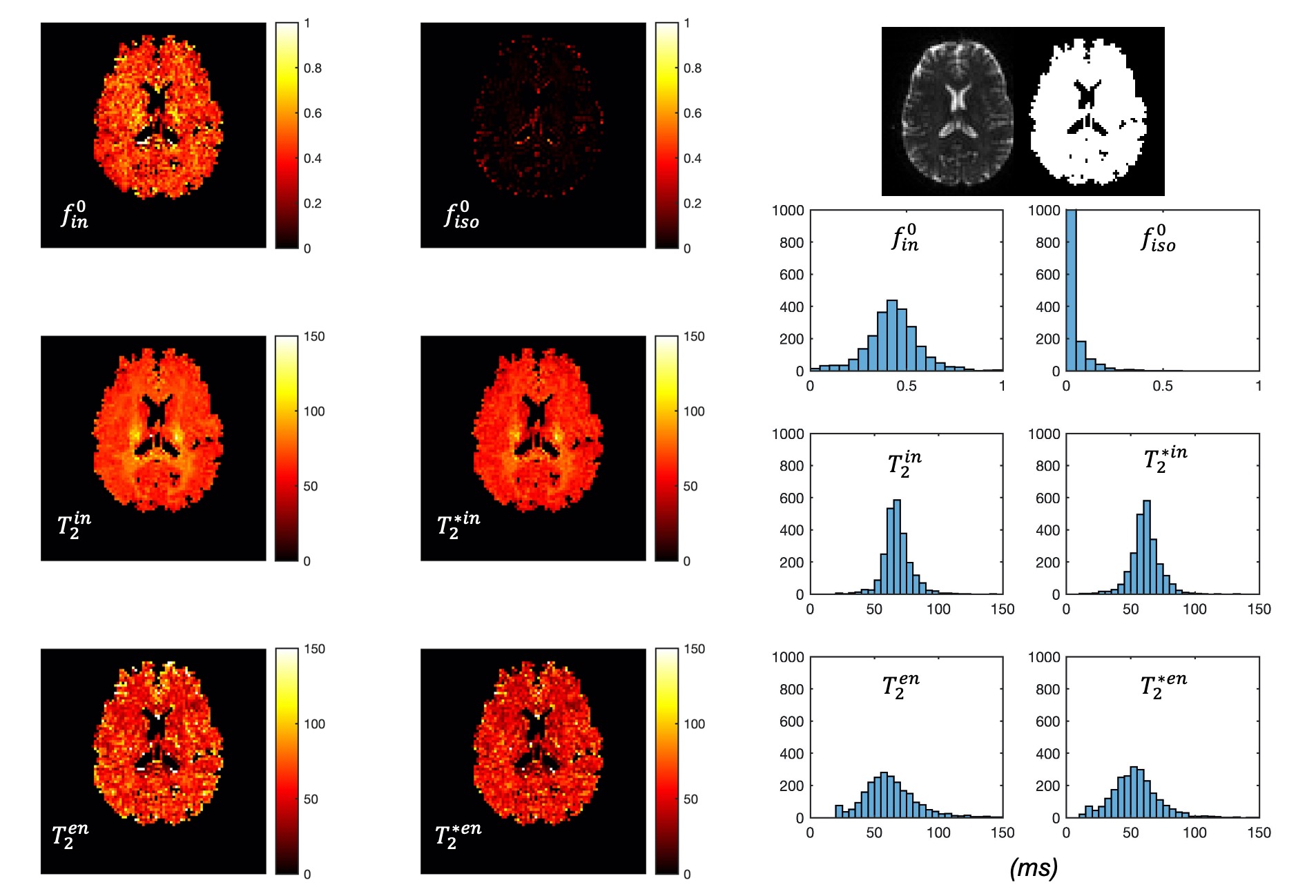}
\caption{Parameter maps and histograms in the brain. Pure CSF voxels are excluded from the maps and histograms. The b=0 image and the inclusion mask are shown (top right).} \label{In-vivo maps}
\end{figure}

\section{Experiments and results}
This section describes the experiments used to evaluate the combined T2-T2*-diffusion imaging technique and the corresponding results. 

\subsection{\textit{\textbf{in vivo}} Study}

\subsubsection{Design} To demonstrate the proposed model with \textit{in vivo} data, a dataset was collected from a healthy subject on a Siemens Prisma scanner using a 32-channel head coil. The local ethical committee approved this study, and a written informed consent was obtained from the participant. A multi-shell diffusion-PEPTIDE imaging protocol is implemented with the following imaging parameters: 4 repetitions at b=0, 20 diffusion gradient directions at b=700 $s/mm^2$ and 30 diffusion gradient directions at b=2000 $s/mm^2$; the TE range is from 84 ms to 131 ms with $\mbox{TE}_{SE} = 108$ ms (48 time points separated by an echo-spacing of 1ms); TR=3500 ms, resolution = 2.5x2.5x3.0 $mm^3$. The dataset hence contains 54×48 volumes, and the total acquisition time is about 32 mins.

\begin{figure}[ht]
\includegraphics[width=\textwidth]{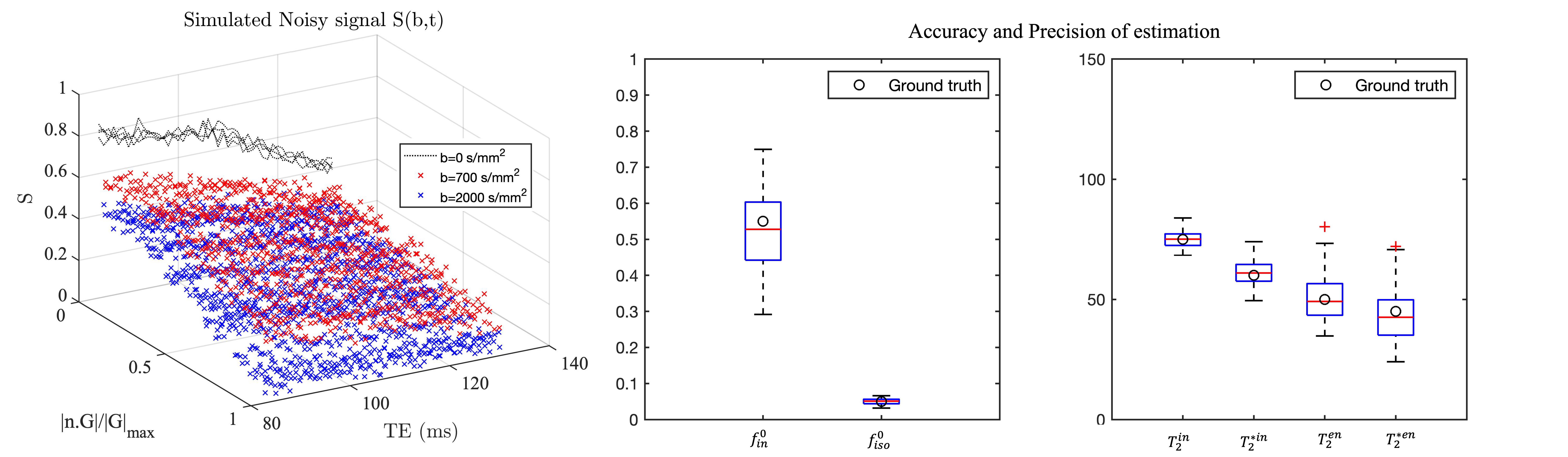}
\caption{Demonstration of simulated noisy signal, and boxplot distributions of estimated parameters from the noisy simulations.} \label{Simulation_estimation_accuracy}
\end{figure}

\subsubsection{Results} Measured and model-predicted diffusion-PEPTIDE signals from a typical white matter (WM) voxel are shown in Fig.~\ref{In-vivo signal}, which agree well in terms of TE and diffusion weighting dependence. Estimated parameter maps are shown in Fig.~\ref{In-vivo maps}. $T_2^{in}$ is higher than $T_2^{en}$ in most WM voxels, which agrees with previous findings~\cite{Gong2020MTE-NODDI:Times,Veraart2018TETimes}; $T_2^{*in}$ and $T_2^{*en}$ estimated from our method in WM also agree well with the range found in a previous WM T2*-diffusion study~\cite{Kleban2020StrongBrain}.

\subsection{Simulation Study}

\subsubsection{Design} To assess whether the method can fit the parameters accurately and precisely, diffusion-PEPTIDE signals are simulated from the forward model with added Rician noise (100 random realisations). The same imaging parameters as \textit{in vivo} data are used, and tissue parameters are chosen to be the estimated values from a representative WM voxel in the genu of the corpus callosum, with $f_{in}^0=0.55$, $f_{iso}^0=0.05$, $T_2^{in}=75$ ms, $T_2^{*in}=60$ ms, $T_2^{en}=50$ ms, and $T_2^{*en}=45$ ms, which are also in agreement with previous findings~\cite{Gong2020MTE-NODDI:Times,Kleban2020StrongBrain}. The level of added noise achieves a SNR of 30 for $S(b=0,\mbox{TE}=84\mbox{ms})$, which matches the estimated value for the \textit{in vivo} WM voxel with multiple b=0 measurements.

\subsubsection{Results} Distributions of estimated parameters from simulated data (Fig.~\ref{Simulation_estimation_accuracy}) suggest parameters of the typical case can be estimated correctly with minimal bias, and the estimation precision for $T_2^{in}$ and $T_2^{*in}$ is much higher than $T_2^{en}$ and $T_2^{*en}$. This is likely because the $T_2^{in}$ and $T_2^{*in}$ is higher than $T_2^{en}$ and $T_2^{*en}$, hence higher effective SNR for the intra-neurite signals.

\section{Discussion and Conclusion}
In summary, this study proposes a combined T2-T2*-diffusion imaging technique to simultaneously derive the non-relaxation-weighted compartment fractions, compartment-specific T2 and T2* relaxation times. While existing methods can provide either compartmental T2 or T2* depending on acquisition and analysis methods used, we achieve simultaneously mapping of compartmental T2 and T2* by proposing a modelling and fitting method making use of diffusion-PEPTIDE acquisition. Experiments are conducted with both \textit{in vivo} and simulation data, where the model utility is demonstrated with \textit{in vivo} study, and the accuracy and precision of estimation is demonstrated with simulation. 

The compartmental T2 and T2* values estimated from our \textit{in vivo} data correspond well with existing studies estimating either T2 or T2*. For intra/extra-neurite T2 values, our results are largely consistent with those obtained from MTE-NODDI~\cite{Gong2020MTE-NODDI:Times} or TEdDI~\cite{Veraart2018TETimes} method. While these previous methods can map intra- and extra-neurite T2 over the whole brain, compartmental T2* estimation has only been achieved in one previous study~\cite{Kleban2020StrongBrain}, limited to a small number of WM tracts with known fibre orientations and minimal orientation dispersion, and relying on an ultra-high gradient system. To the best of our knowledge, the present study demonstrates a technique that for the first time can map compartment-specific T2-T2*-diffusion simultaneously over the whole brain.

Our simulation results demonstrated the parameters can be estimated accurately with minimal bias. While we try to match the SNR level in the simulation data with \textit{in vivo} data, we found that the effective SNR in simulation is much lower than \textit{in vivo} data as suggested in Fig.~\ref{In-vivo signal}. This is likely because noise is highly correlated in \textit{in vivo} data because of the high GRAPPA kernel used for image reconstruction. In contrast, simulated noises are completely independent.

The multi-stage fitting was developed because a brute-force estimation will require a reliable starting point to avoid local minima, which can be numerous in such a high-dimensional parameter space. One limitation of the proposed multi-stage fitting is the prolonged computation time, as it requires fitting a NODDI model for as many times as the number of TEs. This could be mitigated by deep-learning-based fitting, which will be a future direction of research.

In conclusion, the proposed relaxometry-diffusion combined modelling with diffusion-PEPTIDE imaging enables intra-/extra-neurite T2 and T2* mapping simultaneously in the brain, providing new information about tissue microstructure.  Future work will explore the applications of this method to neuroimaging studies where T2, T2* and diffusion properties reflect complementary information about tissue alterations, such as in the disease progression of Parkinson's disease and multiple sclerosis. 

%
%
%
%
%
\bibliographystyle{splncs04}
\bibliography{references}

\end{document}